# SHORTEST PATHS SEARCH METHOD BASED ON THE PROJECTIVE DESCRIPTION OF UNWEIGHTED MIXED GRAPHS

**V.A. Melent'ev**

*A.V. Rzhanov Institute of Semiconductor Physics, Siberian Branch of the Russian Academy of Sciences*
*630090, Novosibirsk, 13 Lavrentiev Ave., Russia*

**Keywords:** mixed graphs, graph projections, shortest paths, asymptotic complexity

**Abstract:** The method is based on the preliminary transformation of the traditionally used matrices or adjacency lists in the graph theory into refined projections free from redundant information, and their subsequent use in constructing shortest paths. Unlike adjacency matrices and lists based on enumerating binary adjacency relations, the refined projection is based on enumerating more complex relations: simple paths from a given graph vertex that are shortest. The preliminary acquisition of such projections reduces the algorithmic complexity of applications using them and improves their volumetric and real-time characteristics to linear ones for a pair of vertices. The class of graphs considered is extended to mixed graphs.

УДК: 004.738.8 + 519.178

# МЕТОД ПОИСКА КРАТЧАЙШИХ ПУТЕЙ, ОСНОВАННЫЙ НА ПРОЕКТИВНОМ ОПИСАНИИ НЕВЗВЕШЕННЫХ СМЕШАННЫХ ГРАФОВ

**В.А. Мелентьев**

*Институт физики полупроводников им. А.В. Ржанова СО РАН*
*Россия, 630090, Новосибирск, пр. ак. Лаврентьева, 13*

**Аннотация:** Метод основан на предварительном преобразовании традиционно используемых в исследовании графов матриц или списков смежности в очищенные от избыточной информации (рафинированные) проекции с последующим их использованием в построении кратчайших путей. В отличие от матриц и списков смежности, основанных на перечислении бинарных отношений смежности, рафинированная проекция основана на перечислении более сложных отношений: простых путей из заданной вершины графа, являющихся кратчайшими. Предварительное получение таких проекций сокращает алгоритмическую сложность использующих их приложений и улучшает их объемные и реальновременные характеристики до линейной для пары вершин. Класс рассмотренных при этом графов расширен до смешанных.

# 1. Введение

Проблема поиска кратчайших путей занимает центральное место в самых разных областях, начиная от компьютерных наук и заканчивая социальными и транспортными сетями. Из-за постоянно растущей сложности современных сетей эффективность поиска кратчайших путей играет определяющую роль [1].

Исторический экскурс в данной области демонстрирует эволюцию алгоритмов — от классических (Дейкстры [2], Беллмана-Форда [3], А* [4] и Флойда-Уоршалла [5, 6]) до модификации их оптимизацией используемых структур данных [7]. Однако, при масштабировании графов такие подходы сталкиваются с серьёзными временными ограничениями. Это делает проблематичным их применение в приложениях, требующих реальновременной обработки графов с миллионами вершин и рёбер [8], и побуждает исследователей к оптимизации структур данных и поиску других подходов.

Один из таких подходов состоит в замене традиционно используемых описаний графа, основанных на бинарных отношениях смежности вершин, проективным его описанием — системой проекций, использующих более крупные, $n$-арные отношения достижимости остальных вершин графа из одной вершины, определяющей ракурс проекции [9-12]. А так как поисковые алгоритмы по сути своей являются комбинаторными, то оперирование более крупными отношениями существенно снижает вариабельность и, соответственно, сложность базирующихся на них алгоритмов [13].

Суть представленного метода заключается в преобразовании традиционных смежностных описаний графа в проекции, каждая из которых содержит информацию, достаточную для построения кратчайшего пути из ракурсной вершины в остальные вершины графа. В данной работе мы ограничимся рассмотрением SSSP проблемы (Single-Source Shortest Paths), состоящей в поиске всех кратчайших путей из одной вершины. Однако представленной здесь информации достаточно для поиска путей в других постановках, таких как "для пары вершин", "для всех пар вершин" и т.п.

# 2. Проекция невзвешенного смешанного графа

## 2.1. Общие положения

Формальное определение проекции графа дадим в виде последовательности шагов, необходимых для её построения, и продемонстрируем это на примере изображенного на рис. 1 графа и соответствующей ему матрицы смежности (таблица 1):

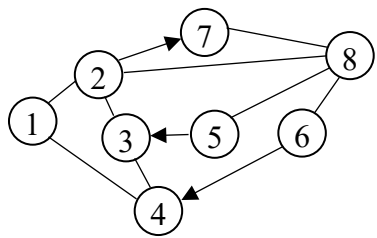

**Рис 1.** Простой смешанный ненагруженный граф

**Таблица 1.** Матрица смежности $A$ графа $G(V, E)$, изображенного на рис. 1.

| $v_i$ \ $v_j$ | 1 | 2 | 3 | 4 | 5 | 6 | 7 | 8 |
|---|---|---|---|---|---|---|---|---|
| 1 | # | 1 | ∅ | 1 | ∅ | ∅ | ∅ | ∅ |
| 2 | 1 | # | 1 | ∅ | ∅ | ∅ | 1 | 1 |
| 3 | ∅ | 1 | # | 1 | ∅ | ∅ | ∅ | ∅ |
| 4 | 1 | ∅ | 1 | # | ∅ | ∅ | ∅ | ∅ |
| 5 | ∅ | ∅ | 1 | ∅ | # | ∅ | ∅ | 1 |
| 6 | ∅ | ∅ | ∅ | 1 | ∅ | # | ∅ | 1 |
| 7 | ∅ | ∅ | ∅ | ∅ | ∅ | ∅ | # | 1 |
| 8 | ∅ | 1 | ∅ | ∅ | 1 | 1 | 1 | # |

Символ "∅" (пусто) в элементе (ячейке) $a_{ij}$ матрицы $A$ означает отсутствие дуги или ребра из вершины $v_i$ $i$-й строки, в вершину $v$ $j$-го столбца, тогда как $a_{ij} = 1$ говорит о наличии дуги $(v_i, v_j)$ из $v_i$, в $v_j$, если $a_{ji} = ∅$, или ребра $(v_i, v_j)$, если $a_{ij} = a_{ji} = 1$.

*Проекция* $P(u)$ графа $G(V, E)$ представляет собой $k$-уровневую конструкцию, на нулевом уровне которой расположена ракурсная вершина $u \in V$. Порожденное этой вершиной подмножество $V_1 \subset V$ смежных из $u$ вершин размещено на первом уровне и содержит все вершины ее окружения $V_1 = \mathcal{N}(u) = \{(v_{1,1})_w, (v_{1,2})_w, \ldots, (v_{1,|\mathcal{N}(u)|})_w\}$.

Каждая вершина первого уровня $v_{1,x} \in V_1$ порождает на следующем уровне подмножество $\{\mathcal{N}(v_{1,1}) \setminus \{u\}\}$ смежных с ней вершин, исключая предшествующую вершинам $v_{1,x} \in V_1)$ вершину $u$. Совокупность этих подмножеств определяет все вершины 2-го уровня — $V_2 = \{\mathcal{N}(v_{1,1}) \setminus \{u\}, \mathcal{N}(v_{1,2}) \setminus \{u\}, \ldots, \mathcal{N}(v_{1,|\mathcal{N}(u)|}) \setminus \{u\}\}$.

Каждый $i$-й ($i > 1$) уровень проекции представляет собой совокупность окружений вершин ($i - 1$)-го уровня, исключающих все предшествующие им вершины, вплоть до $u$. Например, для вершины $v_{(i-1),x}$ на $i$-м уровне создается подмножество смежных ей вершин $V_i(v_{(i-1),x})$ без предшествующих ей на пути из $u$ в $v_{(i-1),x}$ вершин: $V_i(v_{(i-1),x}) = \mathcal{N}(v_{(i-1),x}) \setminus (u$-$v_{(i-1),x})$. Для вершины 1 изображенного на рис. 1 графа:

$$P(1) = 1^{(2^{(3^{(4)},7^{(8)},8^{(5,6,7)})}, 4^{(3^{(2)})})}.$$

Число порожденных на $i$-м уровне подмножеств равно числу экземпляров вершин ($i - 1$)-го уровня. Некоторые из этих подмножеств могут быть пустыми, если $\mathcal{N}(v_{(i-1),x}) \setminus (u$-$v_{(i-1),x}) = \emptyset$ [13, 14].

Число уровней в проекциях графа $G(V, E)$ зависит от поставленной задачи: здесь используются полновершинные проекции графа, покрывающие все вершины графа. Показанная выше 3-уровневая проекция $P(1)$ является полновершинной, так как множество ее вершин покрывает множество вершин $V$ графа $G(V, E)$, и вершины 1 — = 3.

Проекция $P(u)$ графа $G(V, E)$, построенная до уровня $k$, равного эксцентриситету $\varepsilon(u)$, определяет все простые пути из ракурсной вершины $u \in V$ в каждую вершину мультимножества $V'(k) = \bigcup_{i=1}^{k} V_i'$, $V_i' \subseteq V_i$, $|V_i'| \geq |V_i|$ вершин всех уровней, расстояния до которых, измеряемые числом ребер и дуг в соответствующих путях, не превышают $k$.

Однако, такое достаточно полное описание не всегда выгодно: конкретизация поставленных перед исследователем целей позволяет отбросить некоторые детали в формальном описании исследуемого объекта и упростить достижение этих целей. В приложении к рассматриваемой здесь проблеме поиска, такая проекция графа содержит информацию обо всех путях, даже о тех, которые заведомо не могут быть кратчайшими. Поэтому далее мы сформулируем и обоснуем утверждения, позволяющие исключить из проекции избыточную для задачи поиска кратчайших путей информацию.

## 2.2. Рафинирование проекции взвешенного графа для задачи поиска кратчайших путей.

В основе сформулированной ниже леммы лежит принцип оптимальности Беллмана-Дейкстры [15, 16], согласно которому любые участки кратчайшего пути в положительно взвешенных графах тоже являются кратчайшими.

*Лемма 1. Если k-уровневая проекция P(u) содержит несколько экземпляров вершины v, то кратчайшими будут только пути, которые проходят через экземпляры v, находящиеся на самом нижнем уровне проекции.*

Эта лемма является прямым следствием упомянутого выше принципа Беллмана-Дейкстры. В соответствии с ней из исходной проекции $P(u)$ может быть исключена избыточная информация обо всех заведомо не кратчайших путях в экземпляры вершин. Кроме того, надстраивание проекции из таких экземпляров также может быть прекращено, т. к. пути в такие экземпляры обладают длиной, большей в сравнении с другими путями, определяемыми выстраиваемой проекцией, а следовательно, не являются кратчайшими ни эти пути, ни возможные их продолжения.

Покажем применение леммы 1 к проекции $P(4)$ демонстрационного графа. В результате получена рафинированная проекция $P'(4)$, из которой экземпляры вершин, не удовлетворяющие условиям леммы 1, были исключены. Для лучшего визуального сравнения исходной $P(4)$ и полученной $P(4)'$ проекций, исключенные экземпляры вершин в последней скрыты (помечены серым цветом).

$$P(4) = 4^{(1^{(2^{(3^0, 7^{(8)}, 8^{(5,6,7)})}, 3^{(2^{(1^0, 7^{(8)}, 8^{(5,6,7)})})})}}, \rightarrow$$

$$\rightarrow P'(4) = 4^{(1^{(2^{(3^0, 7^{(8)}, 8^{(5,6,7)})}, 3^{(2^{(1^0, 7^{(8)}, 8^{(5,6,7)})})})})}, \rightarrow P'(4) = 4^{(1^{(2^{(7^0, 8^{(5,6)})}, 3^{(2^{(7^0, 8^{(5,6)})})})})}.$$

Рафинированная проекция $P'(u)$ может быть получена непосредственно из матрицы смежности, минуя этап получения $P(u)$, если в процедуру ее построения включить проверку на условия, определенные леммой 1. Хотя удаление из $P(u)$ вершин, входящих в составы заведомо не кратчайших путей, может привести к увеличению числа уровней в проекции $P'(u)$, но в ней при этом исключена репликация вершин, удаленных от $u$ на превышающее кратчайшее расстояние. Таким образом, проекция $P'(u)$ определяет только кратчайшие пути из ракурсной вершины $u$ во все остальные вершины графа.

## 2.3. Построение рафинированной проекции $P'(u)$ и поиск кратчайших путей из вершины $u$.

Конечной целью построения проекции $P'(u)$ является получение компактной структуры данных, из которой исключены все пути, не являющиеся кратчайшими.

Оформим $P'(u)$ в виде строки, обозначим ее $B(u)$, номер ячейки $j$ в ней соответствует вершине $v_j$ простого пути из $u$ в $v_j$, а содержимое $b_j$ — номер вершины, предшествующей вершине $v_j$ на пути из $u$ в $v_j$. Дополнительно используем следующие обозначения:

$A$ — матрица смежности графа, $a_{i,j}$ — содержимое ячейки $(i,j)$ пересечения $i$-й строки и $j$-го столбца матрицы $A$, $A(i)$ — множество вершин $i$-й строки матрицы $A$, смежных из вершины $i$, $A(x) \equiv \mathcal{N}(x)$ — окружение вершины $x$;

$R_{u\text{-}v} = \{u, \ldots, v\}$ — множество вершин, составляющих простой путь из $u$ в $v$ и запрещенных для продолжения этого пути из вершины $v$;

$V_A$ — множество вершин, кратчайшие пути в которые известны, множество блокируемых для чтения столбцов матрицы смежности $A$;

$V_k$ — мультимножество вершин $k$-го уровня проекции $P'(u)$, $V_k(x)$ — множество вершин $k$-го уровня $P'(u)$, порождаемых вершиной $x$ $(k - 1)$-го уровня;

Процесс получения однострочной рафинированной проекции $P'(u)$ осуществляем пошагово, где каждый шаг $k$ соответствует текущему уровню ее построения:

1) $k:=1$. Множество $V_1$ порожденных на 1-м уровне вершин равно окружению $\mathcal{N}(u)$: $V_1 := \mathcal{N}(u) = A(u)$, поэтому $b_{x \in V_1} := u$. Для каждой вершины $x \in V_1$ определим множество $R_{u\text{-}x}$ вершин, запрещенных к продолжению через нее: $R_{u\text{-}x} \mid x \in V_1 := \{u, x\}$. Формируем множество $V_A$ вершин, кратчайшие пути в которые найдены (включаем в него и ракурсную вершину $u$ проекции): $V_A := \{u\} \cup \mathcal{N}(u)$. В соответствии с леммой 1 блокируем доступ к чтению столбцов матрицы смежности $A$, соответствующих вершинам из $V_A$. Если число $|V_A|$ найденных кратчайших путей (заблокированных столбцов) совпадает с числом $|V|$ вершин в графе, то, то проекция $P'(u)$ создана, иначе переходим к п.2.

2) Наращиваем номер текущего уровня проекции: $k := k + 1$. Для каждой вершины предыдущего $k-1$-го уровня определяем подмножество вершин, порождаемых на текущем $k$-м уровне проекции: $V_k(x \in V_{k-1}) := \mathcal{N}(x) \setminus V_A \setminus R_{u\text{-}x}$. При этом окружения $\mathcal{N}(x \in V_{k-1}(x))$ определяются из соответствующих вершинам $x \in V_{k-1}(x)$ строк модифицированной на предшествующих шагах матрицы смежности $A$ (с заблокированными для чтения столбцами, соответствующими найденным ранее кратчайшим путям): $\mathcal{N}(x \in V_{k-1}(x)) := \mathcal{N}(x) \setminus V_A$. В ячейки $B$, соответствующие вершинам $y \in V_k(x \in V_{k-1})$ заносим предшествующие $y$ вершины $x$: $b_y(y \in V_k(x \in V_{k-1})) := x$. Мультимножество вершин текущего уровня получаем объединением этих подмножеств: $V_k := \cup V_k(x \mid x \in V_{k-1}(x))$. Для каждой вершины $x \in V_k$ текущего уровня $k$ формируем подмножество вершин $R_{u\text{-}x}$, запрещенных к продолжению пути из нее: $R_{u\text{-}x} := R_{u\text{-}x} \cup \{x\} \mid x \in V_k$. Если $|V_A| + |V_k| = |V|$, то сбрасываем блокировку столбцов ($V_A := \emptyset$), т. к. построение проекции завершено, иначе — блокируем для чтения соответствующие $V_k$ столбцы матрицы смежности $A$, модифицируем $V_A := V_A \cup V_k$ и повторяем п.2.

Покажем это на примере проекции $P(4)$ (для наглядности выполняемые шаги нумеруем в соответствии с полученными при этом уровнями):

1) $k := 1$, $V(4) = V_1(4) := A(4) = \{1, 3\}$, $b_1 := 4$, $b_3 := 4$. $R_{4\text{-}1} := \{4, 1\}$, $R_{4\text{-}3} := \{4, 3\}$. $V_A := \{4, 1, 3\}$. Так как $|V_A| < |V|$ ($3 < 8$), условие вершинной полноты не выполнено, то переходим к выполнению п. 2.

2) $k := 2$. Учитывая, что $V_2 = \{V_2(x) \mid x \in V_1\} = \{V_2(x) \mid x \in \{1, 3\}\} = \{V_2(1), V_2(3)\}$, получим $V_2(1) := A(1) \setminus V_A = \{2, 4\} \setminus \{4, 1, 3\} = \{2\}$, $V_2(3) := A(3) \setminus V_A = \{2, 4\} \setminus \{4, 1, 3\} = \{2\}$. $V_2 := \{2\}$, $b_2 = \{1, 3\}$ — это означает, что в вершину 2 из вершины 4 ведут два одинаковых по длине пути: (4, 1, 2) и (4, 3, 2), обозначим их $(4\text{-}2)\_1$ и $(4\text{-}2)\_2$. Тогда $R_{4\text{-}2\_1} = \{4, 1, 2\}$, $R_{4\text{-}2\_2} = \{4, 3, 2\}$. $V_A := \{4, 1, 3\} \cup \{2\} = \{4, 1, 3, 2\}$. Так как $|V_A| < |V|$ ($4 < 8$), условие вершинной полноты не выполнено, возвращаемся в п. 2.

3) $k := 3$. Так как в вершину 2 ведут два пути, то $V_3 = \{V_3(2_1), V_3(2_2)\} = \{(A(2) \setminus V_A) \setminus R_{4\text{-}2\_1}, (A(2) \setminus V_A) \setminus R_{4\text{-}2\_2}\}$. Содержимое 2-й строки матрицы $A$ с учетом заблокированных столбцов $V_A = \{4, 1, 3, 2\}$: $A(2) \setminus V_A = \{1, 3, 7, 8\} \setminus \{4, 1, 3, 2\} = \{7, 8\}$, поэтому мультимножество $V_3 := \{\{7, 8\} \setminus R_{4\text{-}2\_1}, \{7, 8\} \setminus R_{4\text{-}2\_2}\} = \{\{7, 8\}, \{7, 8\}\}$ и $V_A := V_A \cup \{7, 8\} = \{4, 1, 3, 2, 7, 8\}$. Так как $|V_A| < |V|$ ($6 < 8$), то возвращаемся в п. 2.

4) $k := 3 + 1 = 4$. $V_4 = \{V_4(7), V_4(8)\} = \{(A(7) \setminus V_A) \setminus R_{4\text{-}7}, (A(8) \setminus V_A) \setminus R_{4\text{-}8}\} = \{\emptyset, \{5, 6\}\}$, $b_5 = b_6 := 8$. $V_A := V_A \cup \{5, 6\} = \{4, 1, 3, 2, 7, 8\} \cup \{5, 6\} = \{4, 1, 3, 2, 7, 8, 5, 6\}$, $|V_A| = |V|$, вершинная полнота проекции $P'(4)$ достигнута, окончательный ее вид приведен в таблице.

**Таблица 2.** Рафинированная $P'(4)$-проекция исследуемого графа. Здесь $k = \varepsilon(4) = 4$ — последний уровень проекции $P'(4)$, построенной из вершины 4.

| $k$ \ $v_j$ | 1 | 2 | 3 | 4 | 5 | 6 | 7 | 8 |
|---|---|---|---|---|---|---|---|---|
| 4 | 4 | 1, 3 | 4 | # | 8 | 8 | 2 | 2 |

Кратчайшие пути из вершины $u$ в любую из вершин графа определяем последовательной выборкой вершин, предшествующих каждой вершине искомого пути, начиная с конечной. Например, для кратчайшего пути $(4\text{-}5)_\delta$ получим: $b_5 = 8$, $b_8 = 2$, $b_2 = \{1, 3\}$, $b_1 = b_3 = 4$ — найдены два кратчайших пути: $(4\text{-}5)_\delta = \{(4, 1, 2, 8, 5), (4, 3, 2, 8, 5)\}$.

### 2.4. Оценки сложности

Как видно из предыдущего раздела, поиск кратчайших путей включает в себя 2 этапа: этап предварительного преобразования матрицы смежности графа — $T_1$ и этап выборки из найденных на первом этапе путей — $T_2$. Пространственная сложность $S$ первого этапа составляет $S_1 = n^2 = |V^2|$ и второго — $S_2 = n$ — для поиска одного пути или всех путей из одной вершины, или $S_2 = n^2$ — для всех пар вершин в графе.

Рассмотрим асимптотическую сложность $T_1$ получения рафинированной проекции $P'(u)$. Здесь на каждом шаге определяется множество вершин текущего уровня. При этом общее число обработанных вершин не зависит от числа шагов и равно числу вершин в графе, поэтому $T_1 = O(n)$. К примеру, в $K_n$-графе число уровней (шагов) равно единице, а число обработанных вершин равно $n$, в худшем же случае (гамильтонова графа с одним ориентированным ребром) рассмотрению подлежат $n$ уровней с одной вершиной в окружении, т. е. асимптотика $T_1 = O(n)$ также линейна. Решение задачи в ASSP-постановке (ASSP - All-Pairs Shortest Path) потребует построения всех $P'(u \in V)$ проекций графа, и асимптотическая сложность их построения квадратична: $T_1 = O(n^2)$.

В отличие от предварительного, асимптотическая сложность этапа выборки кратчайшего пути из множества уже заданных проекцией $P'(u)$ кратчайших путей определяется удаленностью $d(u, v)$ вершины $v$ от $u$ и не превышает эксцентриситета вершины-источника $\varepsilon(u)$: $T_2(n) = O(\varepsilon(u))$. Эксцентриситет $\varepsilon(u)$ любой вершины $v \in V$ графа $G(V, E)$ не превышает его диаметра $D(G)$, который зависит не только от размера $n = |V|$ графа, но и от плотности и способа соединения вершин: $T_2(n) = O(D(G, n, s))$. Таким образом, суммарная сложность поиска одного пути $T_1(n) + T_2(n) = O(n) + O(D(G, n, s))$. Учитывая $O(D(G, n, s) \leq O(n)$, получим $T_1(n) + T_2(n) = O(n)$, что во-первых, существенно меньше $T = O(n + m)$ алгоритма Дейкстры и, во-вторых, в отличие от него не зависит от плотности графа. А применение в приложениях предварительно найденных проекций $P'(u)$ повысит эффективность поиска еще более: $D(G, n, s) \ll n + m$ и $T_2 \ll T$.

# 3. Заключение

В докладе представлен основанный на описании невзвешенных смешанных графов проекциями метод поиска кратчайших путей, предоставляющий эффективные решения для оптимизации информационных обменов, транспортно-логистических и прочих операций в множестве практических приложений теории графов. Целью дальнейших исследований автора в этой области является распространение метода на взвешенные графы. Мы надеемся, что научные исследования, связанные с проективным описанием графов и с использованием его в теоретических и практических графовых задачах, будут полезны широкому кругу специалистов.

# Список литературы

1. Schrijver A. On the history of the shortest path problem //Documenta Mathematica. 2012. Vol. 17, No. 1. P. 155.
2. Dijkstra E. W. A note on two problems in connexion with graphs://Numerische Mathematik, 1959, No. 1. P. 269.
3. Bellman R. On a routing problem //Quarterly of Applied Mathematics. 1958. Vol. 16, No. 1. P. 87.
4. Hart P. E., Nilsson N. J., Raphael B. A formal basis for the heuristic determination of minimum cost paths //IEEE transactions on Systems Science and Cybernetics. 1968. Vol. 4, No. 2. P. 100.
5. Floyd R. W. Algorithm 97: shortest path //Communications of the ACM. 1962. Vol. 5, No. 6. P. 345.
6. Warshall S. A theorem on boolean matrices //Journal of the ACM (JACM). 1962. Vol. 9, No. 1. P. 11.
7. Kumawat S., Dudeja C., Kumar P. An extensive review of shortest path problem solving algorithms //5th International Conference on Intelligent Computing and Control Systems (ICICCS). – IEEE, 2021. P. 176.
8. Peng W. et al. A fast algorithm to find all-pairs shortest paths in complex networks //Procedia Computer Science. 2012. Vol. 9. P. 557.
9. Мелентьев В.А. Скобочная форма описания графов и ее использование в структурных исследованиях живучих вычислительных систем //Автометрия. 2000. Т. 38, №. 4. С. 36.
10. Melent'ev V.A. The bracket Pattern of a Graph //The 6th International Conference on Pattern Recognition and Image Analysis: New Information Technologies, PRIA-6-2002. 2002. С. 57.
11. Мелентьев В.А. Формальные основы скобочных образов в теории графов //Труды Второй Междунар. конф. Параллельные вычисления и задачи управления PACO. 2004. С. 694.
12. Мелентьев В.А. Актуализация описаний и реконфигурация отказоустойчивых систем //Труды III Международной конференции «Параллельные вычисления и задачи управления» PACO'. 2006. С. 785.
13. Melent'ev V.A. Formal method for the synthesis of optimal topologies of computing systems based on the projective description of graphs //The Journal of Supercomputing. 2022. Vol. 78, No. 12. P. 13915-13941.
14. Zadorozhny A.F., Melent'ev V.A. On the topological compatibility of parallel tasks and computing systems //International Journal of Modern Physics C. 2022. Vol. 33. No. 03. – P. 2250040.
15. Bellman R. On a routing problem //Quarterly of applied mathematics. 1958. Vol. 16, No. 1. P. 87.
16. Dijkstra E.W. A Note on Two Problems in Connexion with Graph //Numerische Mathematik. 1959. Vol. l. P. 269.